\documentclass[aps,prb,showpacs,floatfix,twocolumn,byrevtex,superscriptaddress]{revtex4-1}
\usepackage{amsmath}
\usepackage{amssymb}
\usepackage{amstext}
\usepackage{amsopn}
\usepackage{amsfonts}
\usepackage{amsxtra}
\usepackage[english]{babel}
\usepackage{amsmath}
\usepackage{graphicx}
\usepackage{float}
\usepackage{bm}
\usepackage{multirow}
\usepackage{dcolumn}
\usepackage{hyperref}
\usepackage{CJK}

\newcommand{\icm}{\ensuremath{~\textrm{cm}^{-1}}}% % cm-1
\newcommand{\CPFA}{Ca$_{0.86}$Pr$_{0.14}$Fe$_{2}$As$_{2}$}
\newcommand{\CPFAx}{Ca$_{1-x}$Pr$_{x}$Fe$_{2}$As$_{2}$}
\newcommand{\CFA}{CaFe$_{2}$As$_{2}$}
\newcommand{\CCA}{CaCu$_{2}$As$_{2}$}

\begin{document}

\title{Formation of As-As Bond and Its Effect on Absence of Superconductivity in Collapsed Tetragonal Phase of \CPFA\ : An Optical Spectroscopy Study}
\author{Run Yang}
\author{Congcong Le}
\affiliation{Beijing National Laboratory for Condensed Matter Physics, National Laboratory for Superconductivity, Institute of Physics, Chinese Academy of Sciences, P.O. Box 603, Beijing 100190, China}
\author{Lei Zhang}
\affiliation{High Magnetic Field Laboratory, Chinese Academy of Sciences, Hefei 230031, China}
\author{Bing Xu}
\affiliation{Beijing National Laboratory for Condensed Matter Physics, National Laboratory for Superconductivity, Institute of Physics, Chinese Academy of Sciences, P.O. Box 603, Beijing 100190, China}
\author{Wei Zhang}
\affiliation{Beijing National Laboratory for Condensed Matter Physics, National Laboratory for Superconductivity, Institute of Physics, Chinese Academy of Sciences, P.O. Box 603, Beijing 100190, China}
\affiliation{College of Physics, Optoelectronics and Energy \&\ Collaborative Innovation Center of Suzhou Nano Science and Technology, Soochow University, Suzhou 215006, China}
\author{Kashif Nadeem}
\affiliation{Beijing National Laboratory for Condensed Matter Physics, National Laboratory for Superconductivity, Institute of Physics, Chinese Academy of Sciences, P.O. Box 603, Beijing 100190, China}
\affiliation{Department of Physics, International Islamic University, H-10, Islamabad, Pakistan}
\author{Hong Xiao}
\affiliation{Beijing National Laboratory for Condensed Matter Physics, National Laboratory for Superconductivity, Institute of Physics, Chinese Academy of Sciences, P.O. Box 603, Beijing 100190, China}
\author{Jiangping Hu}
\affiliation{Beijing National Laboratory for Condensed Matter Physics, National Laboratory for Superconductivity, Institute of Physics, Chinese Academy of Sciences, P.O. Box 603, Beijing 100190, China}
\affiliation{Collaborative Innovation Center of Quantum Matter, Beijing 100084, China}
\author{Xianggang Qiu}
\email[]{xgqiu@iphy.ac.cn}
\affiliation{Beijing National Laboratory for Condensed Matter Physics, National Laboratory for Superconductivity, Institute of Physics, Chinese Academy of Sciences, P.O. Box 603, Beijing 100190, China}
\affiliation{Collaborative Innovation Center of Quantum Matter, Beijing 100084, China}

\date{\today}
%%%%%%%%%%%%%%%%%%%%%%%%%%%%%%%%%%%%
%
% Abstract
%

\begin{abstract}
The temperature dependence of in-plane optical conductivity has been investigated for \CPFA\ which shows a structural transition from tetragonal (T) to collapsed tetragonal (cT) phase at $T_{cT}\sim$73 K. Upon entering the cT phase, drastic change characterized by the formation of a mid-infrared peak near 3200\icm(0.4 eV) in the optical conductivity is observed. Analysis of the spectral weight reveals reduced electron correlation after the cT phase transition. Based on the calculated band structure and simulated optical conductivity, we attribute the new feature around 0.4 eV to the formation of interlayer As-As bond. The As-As bond strongly affects the Fe-As hybridizations, and in turn, drastically changes the \CPFA\ into a nonmagnetic Fermi liquid system without bulk superconductivity in the cT phase.
\end{abstract}

%  72.15.-v  Electronic conduction in metals and alloys
%  74.70.-b  SC: Superconducting materials other than cuprates
%  78.20.-e  Optical properties of bulk materials and thin films
%  78.30.-j  Infrared and Raman spectra


\maketitle

%%%%%%%%%%%%%%%%%%%%%%%%%%%%%%%%%%%%%%%%%%%%%%%%%%%%%%%%%%%%%%%%%%%%%%%%%%%%%%%
%
% Introduction

Electron correlation and magnetism are widely believed to be intimately related to the pairing mechanism of unconventional high-temperature superconductor~\cite{Lee}. The iron-pnictides are newly discovered superconductors with $T_c$ as high as 55~K~\cite{Johnston2010}. Similar to the cuprates, superconductivity in iron-pnictides appears in the vicinity of an antiferromagnetic (AFM) phase by doping or pressure. However, the parent compounds of iron-pnictides are bad metals with moderate correlation, and the local magnetic moment in iron-pnictides is much larger than that in the cuprates. The strong magnetism and moderate correlation provide us an alternative material to investigate the interplay among electron correlation, magnetism and superconductivity~\cite{Pag}.

Up to now, there is growing evidence that the covalent Fe-As bond plays an important role in the appearance of superconductivity in iron-pnictides~\cite{Belashchenko2008}. It not only transports the carriers to tune the correlation~\cite{diehl} but also delivers the antiferromagnetic superexchange interactions to form the collinear antiferromagntic order and induces spin fluctuation~\cite{Ma2008}. In addition, first-principle calculations point out that the Fe local moment is strongly sensitive to the Fe-As distance which is influenced by the covalent Fe-As bonding~\cite{Mirbt2003}. Therefore, the Fe-As bonding is one of the key elements for understanding the correlation, magnetism as well as superconductivity in iron-pnictides.

Among the Fe-As based superconductors, \CFA\ is an excellent prototype material to explore the nature of superconducting mechanism due to its strong Fe-As bonding and large structural instability~\cite{Sapa2014,Tom,Gre2013}. At ambient pressure, it crystallizes in the tetragonal structure at room temperature and manifests a strongly first-order, concomitant structural-magnetic phase transition to an orthorhombic phase with the stripe-like antiferromagnetic order at 170~K~\cite{can}. Charge doping and hydrostatic pressure can suppress the AFM order and induce superconductivity in it~\cite{Tori2008}. However, when the pressure exceeds $0.35$~GPa, \CFA\ undergoes a remarkable structure transition from the orthorhombic phase directly into the so-called collapsed tetragonal phase~\cite{Yu}, which can also be stabilized at ambient pressure by introducing chemical pressure~\cite{saha} or post-annealing treatment~\cite{Sapa2014}. This transition is characterized by the formation of interlayer As-As bond and a shrinkage of the $c$ axis by approximately 10\% without breaking any symmetries~\cite{saha}. In contrast to the tetragonal phase at high temperatures, neutron scattering and nuclear magnetic resonance  demonstrated that the local Fe moment~\cite{Ma2013} and the spin fluctuation are quenched in the cT phase\cite{Soh2013}. Recent transport measurements further revealed that following the lattice-collapse transition, Fermi-liquid behavior abruptly recovers along with the disappearance of bulk superconductivity~\cite{kas,Yu}. Very recently, first-principle calculations also show that the hybridizations between Fe $3d$ orbitals and As $4p$ orbitals were also greatly affected by the new formed As-As bond across the phase transition~\cite{Yildirim2009,diehl}. There results suggest that  the vanishing of correlation effect and  magnetism  may be strongly connected to the formation of the As-As bond between adjoint layers.

In this work, we carried out transport, optical spectral measurements  in Pr  doped \CFA\ system. The single crystal \CPFA\ undergoes a first order transition to the cT phase at 73~K.  A drastic spectral change across the phase transition due to a band reconstruction is observed by the optical spectroscopy measurement~\cite{wang}. The spectral weight analysis reveals a reduced correlation and a newly formed mid-infrared peak at 0.4~eV in the cT phase. Comparing with the theoretical results from first-principle calculations, we find that this mid-infrared peak is associated with the formation of interlayer As-As bond, which strongly affects the Fe-As hybridizations and changes the material to a nonmagnetic Fermi-liquid system without bulk superconductivity.

%%%%%%%%%%%%%%%%%%%%%%%%%%%%%%%%%%%%%%%%%%%%%%%%%%%%%%%%%%%%%%%%%%%%%%%%%%%%%%%
%
% Experiments
%
High quality single crystals \CPFAx with nominal concentration of $x=$0.14 were synthesized by the FeAs self-flux method and the typical size was about 10$\times$10$\times$0.1~mm$^3$ ~\cite{ron}. The composition $x$ determined by Inductive Coupled Plasma Emission Spectrometer (ICP) was about $0.135\pm 0.05$ . The lattice structural was detected by X-ray diffraction (XRD) (Bruker D8 Advance instrument) using Cu K$\alpha$ ($\lambda$= 0.154 nm) radiation at various temperatures upon warming. Resistivity measurement was carried out on Quantum Design Physical Property Measurement System (PPMS). Magnetization was measured using a Quantum Design superconducting quantum interference device (SQUID).
The reflectivity from the cleaved surface has been measured at a near-normal angle of incidence on a Fourier transform infrared spectrommeter (Bruker 80v) for light polarized in the $ab$ planes using an $in~situ$ evaporation technique.~\cite{Chri} Data from 40 to 15\,000~\icm\ were collected at 8 different temperatures from 15~K to 300~K on an ARS-Helitran cryostat. The reflectivity in the visible and UV range (10\,000-40\,000~\icm) at room temperature was taken with an Avaspec 2048$\times$14 optical fiber spectrometer. The optical conductivity has been determined from a Kramers-Kronig analysis of reflectivity $R(\omega)$ over the entire frequency range. A Hagen-Rubens relation ($R=1-A\sqrt{\omega}$) is used for low-frequency extrapolation. Above the highest-measured frequency (40\,000~\icm), $R(\omega)$ is assumed to be constant up to 40~eV, above which a free-electron response ($\omega^{-4}$) is used~\cite{Dai}.

%%%%%%%%%%%%%%%%%%%%%%%%%%%%%%%%%%%%%%%%%%%%%%%%%%%%%%%%%%%%%%%%%%%%%%%%%%%%%%%
%
% Results
%

\begin{figure}[tb]
\includegraphics[width=0.65\columnwidth]{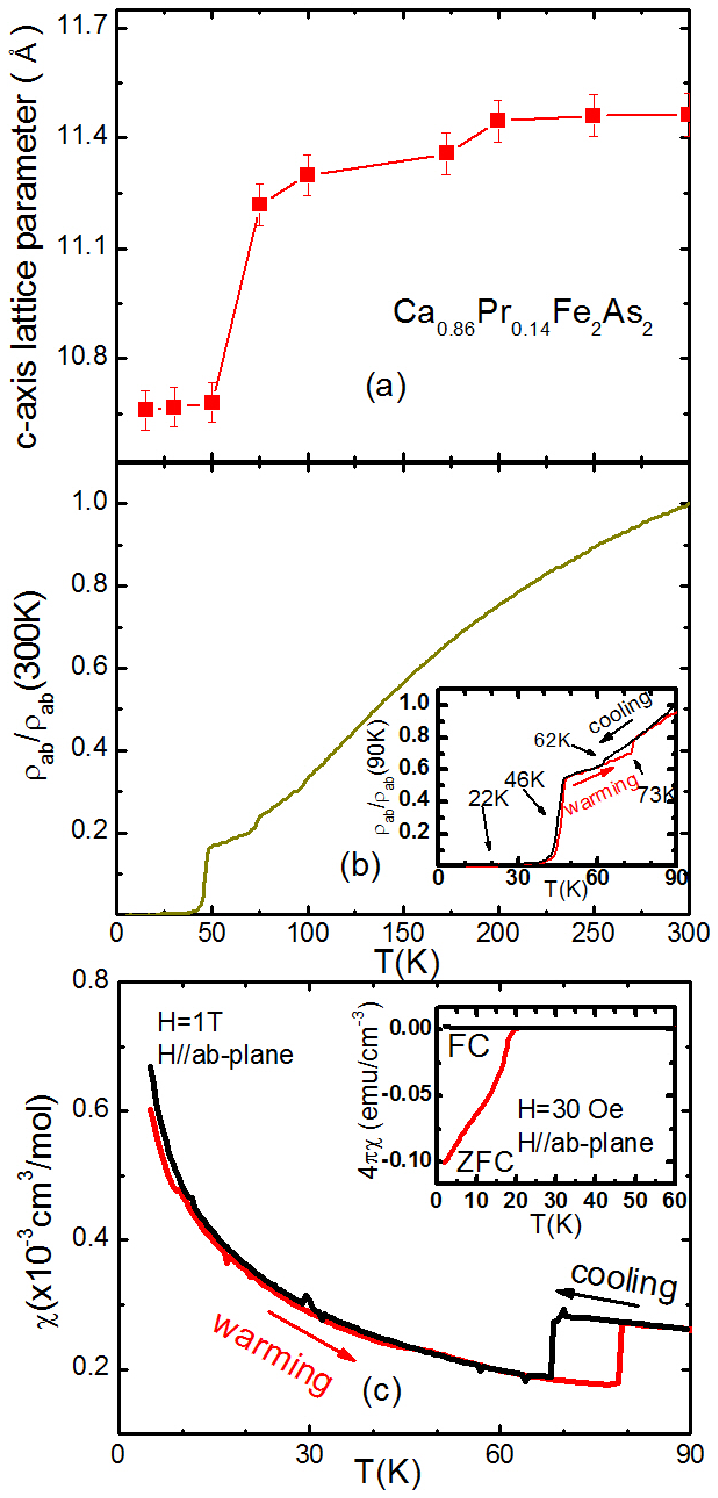}
\caption{ (color online)(a)Temperature dependence of lattice parameter c at various temperatures upon warming. (b)Temperature dependence of normalized resistivity of \CPFA. The inset plots an enlarged low temperature data. (b)Temperature dependence of magnetic susceptibility  of \CPFA with $H\parallel ab$ plane at $1$~T.The inset shows the zero-field-cooled and field-cooled at 30~Oe.}
\label{RT}
\end{figure}

In Fig.~\ref{RT}(a) we calculated the c-axis parameter of \CPFA\ from the XRD data, a great shrinkage happened below 75~K indicates the lattice collapse transition~\cite{saha}. Correspondingly, a sudden drop is observed near 73~K in its in-plane resistivity (show in the main panel of Fig.~\ref{RT}(b)). The thermal hysteretic loop in the inset of Fig.~\ref{RT}(b) indicates the first order nature of this transition. From the dc magnetic susceptibility (shown in Fig.~\ref{RT}(c)), a kink combined with a thermal hysteresis can also be found near 73~K. Even though obvious drops at 46~K and zero-resistance behavior at 21~K are observed in the resistance curve~\cite{Lv2011}, the shielding volume fraction estimated from the field-cooling curve and zero-field-cooling curve in the inset of Fig.~\ref{RT}(c) is about 10\%, which is much smaller than that expected for a bulk superconductor~\cite{saha}. Furthermore, the superconductivity is easily to be suppressed by a magnetic field of 1~T, suggesting its interfacial or filamentary nature~\cite{Gofry2014}.

\begin{figure}[tb]
\includegraphics[width=0.8\columnwidth]{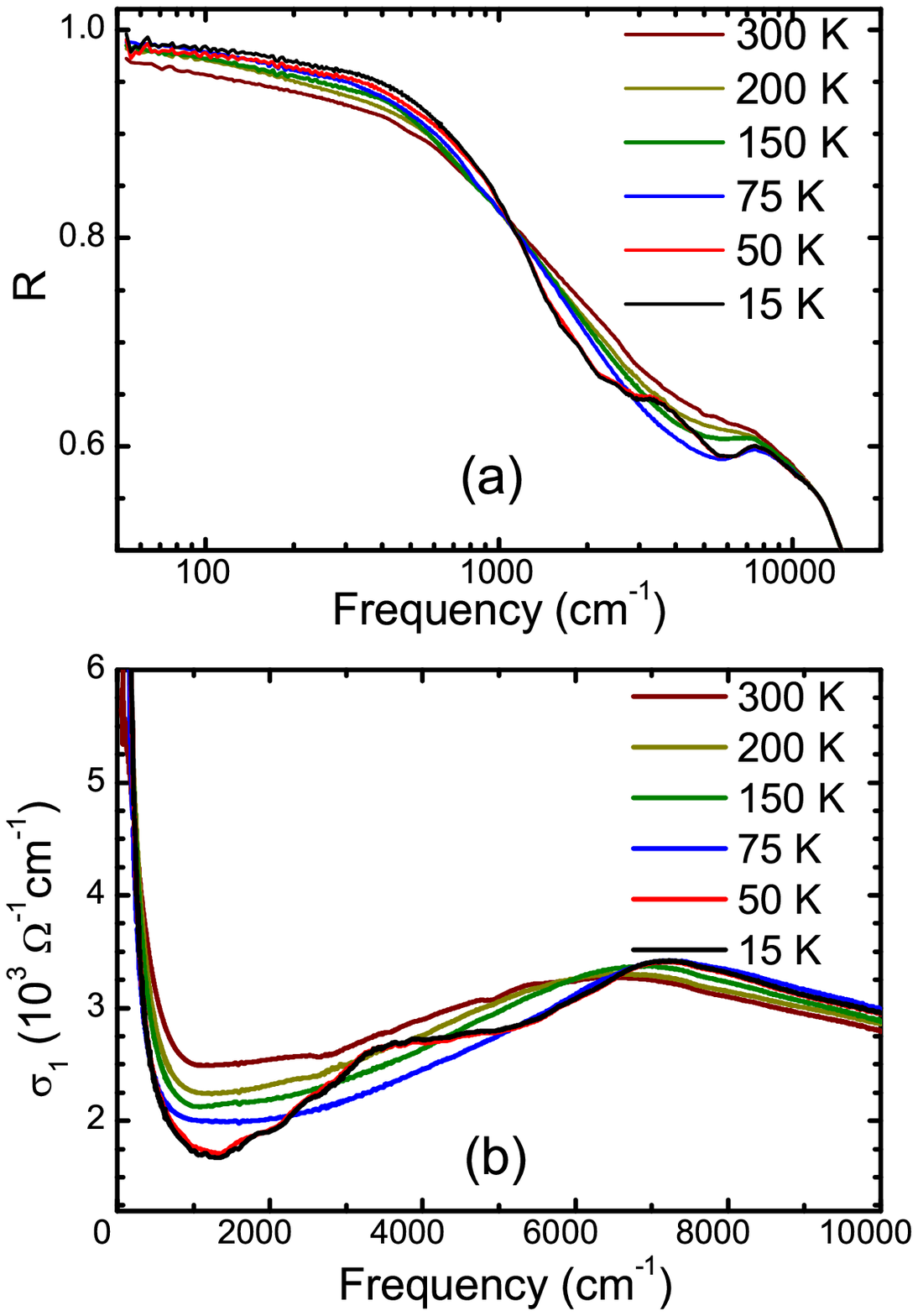}
\caption{ (color online)Reflectivity (a) and optical conductivity (b) of \CPFA\  at various temperatures }
\label{optical}
\end{figure}

The measured in-plane reflectivity $R(\omega)$ and the real part of the in-plane optical conductivity $\sigma_1(\omega)$ are presented in Fig.~\ref{optical}, respectively, for selected temperatures above and below $T_{cT}$. The reflectivity shows a typical metallic response, approaching unity at low frequencies ($< 1\,000~\icm$) and increasing upon cooling. Upon entering the cT phase,  the $R(\omega)$ in the mid-infrared range (1\,000-3\,000~\icm) is greatly suppressed. As a consequence, the reflectivity edges at about 1\,000~\icm become sharper, suggesting a suppression of the carrier scattering. As suggested by Basov et al. and Pratt et al.~\cite{Basov2005,Pratt}, the scattering could be caused by spin fluctuation, shaper reflectivity edges also indicate diminishing spin fluctuation in the cT phase. The drastic change could also be seen from the optical conductivity data in Fig.~\ref{optical}(b), below $T_{cT}$, an abrupt decrease in the conductivity between 1\,000~\icm\ and 2\,500~\icm and a peak at 3\,500~\icm\ ($\sim$0.4~eV) are observed (see Appendix ~\ref{APPENDIX A} for more detail). Similar behaviors have also been observed in $R(\omega,~T)$ and $\sigma_1(\omega,~T)$ data of P-doped \CFA\ , which have been attributed to an abrupt band reconstruction across the cT transition ~\cite{wang}.

%
% Figure 1
%
%
% Figure 2
%
% Figure 3
%
\begin{figure}[tb]
\includegraphics[width=1.0\columnwidth]{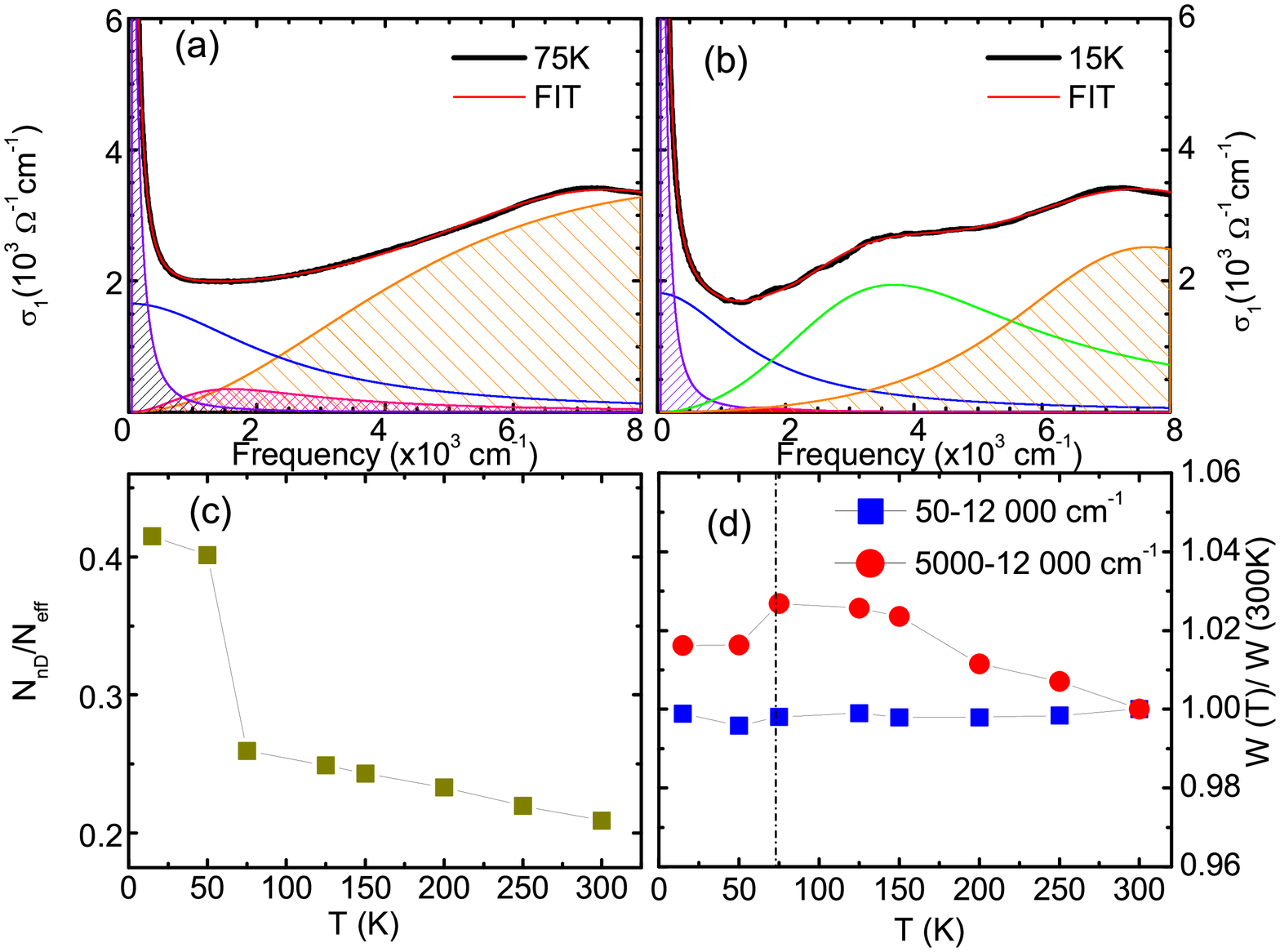}
\caption{(color on line) (a) (b) Optical conductivity of \CPFA\ at 75~K (T phase) and 15~K (cT phase) (thick black lines), fitting with the Drude-Lorentz model (thin red lines) and its decomposition into individual Drude and Lorentz terms. (c) Weights of the narrow Drude component ($N_{nD}$) and the total weights of the narrow and broad Drude components ($N_{eff}$) for various temperatures. (d) Temperature dependence of the spectral weight, $W^{\omega_b}_{\omega_a}~=~\int^{\omega_b}_{\omega_a}\sigma_1(\omega)d\omega$ between different lower and upper cutoff frequencies. The vertical dashed line denotes $T_{cT}$. }
\label{fit}
\end{figure}
%%%%%%%%%%%%%%%%%%%%%%%%%%%%%%%%%%%%%%%%%%%%%%%%%%%%%%%%%%%%%%%%%%%%%%%%%%%%%%%
%
% Discussion
%
%
To quantitatively analyze the optical data of \CPFA, we fit the $\sigma_1(\omega,~T)$ with a simple Drude-Lorentz mode ~\cite{Wu,Dai}:
\begin{equation}
\label{DrudeLorentz}
\epsilon(\omega)=\epsilon_{\infty}-\sum_{i}\frac{\Omega^{2}_{p,i}}{\omega^{2}+\frac{i\omega}{\tau_{i}}}+\sum_{j}\frac{\Omega^{2}_{j}}{\omega^2_{j}-\omega^2-\frac{i\omega}{\tau_{i}}},
\end{equation}
where $\epsilon_{\infty}$ is the real part of the dielectric function at high frequencies, the second term corresponds to the Drude response characterized by a plasma frequency $\Omega_{p,i}^{2}=4\pi ne^{2}/m^{*}$, with $n$ a carrier concentration and $m^{*}$ an effective mass, and $1/\tau_i$ the scattering rate. The third term is a sum of Lorentz oscillators characterized by a resonance frequency $\omega_{j}$, a linewidth $\gamma_{j}$, and an oscillator strength $\Omega_{j}$. The Drude term accounts for the free carrier (intraband) response, while the Lorentz contributions represent the interband excitations~\cite{marsik,cal}. The fitting results shown in Fig.~\ref{fit}(a)(b) can well reproduce the experimental results regardless of the detail band structure.

From the fitting results (Fig.~\ref{fit}(a)(b) see more results in the supplementary material), we find that the optical conductivity at low frequency ($<2\,000$\icm) is dominated by two Drude (intraband) responses, a broad one and a narrow one, reflecting the multi-orbital nature of the iron pnictides (see Appendix ~\ref{APPENDIX B} for a detail fitting results).
The narrow Drude item represents the response of the coherent carriers, corresponding to the electron pockets. The other Drude component with an extremely broad width of 2\,300~\icm\ or more is originated from a highly incoherent scattering and is dominated by the hole pockets at the center of Brillouin zone.~\cite{Wu,Dai} It has been pointed out that the fraction of coherent Drude weight $N_{nD}$ ($N_{nD}=\frac{2m_{0}V}{\pi e^{2}}\int^{\infty}_{0}\sigma_{nD}(\omega)~d\omega$ in which $m_{0}$, $V$, $\sigma_{nD}$ denote the free electron mass, the cell volume and the narrow Drude component, respectively.) at low frequency represents the degree of coherence. Here we define $N_{eff}$ as the sum of the weight of the narrow and broad Drude components in order to estimate the low-energy intraband response~\cite{naka}. Since the electronic correlation has been regarded as a source of incoherence in iron arsenides, $N_{nD} / N_{eff}$ could also be a measure of the strength of electronic correlations~\cite{naka}. The data of $N_{nD} / N_{eff}$ in Fig. 3(c) shows an abrupt enhancement of the degree of coherence across the T to cT phase transition, suggesting a weakening of electronic correlation.

Then we focus on the optical conductivity $\sigma_1(\omega)$ at high frequencies (above 5\,000\icm). It has been reported that in iron-pnictides the spectral weight at low frequency would transfer to high energy area ($>~0.5$~eV) as temperature goes down. This behavior is widely attibuted to the Hund's rule coupling effect~\cite{WangNL,Schafgans2012}, which can localize and polarize the itinerant electrons to enhance the correlation and local moment. To investigate the anomalous spectral weight transfer, we have calculated and normalized the spectral weight between different lower and upper cutoff frequencies. From the results shown in Fig.~\ref{fit}(d), it can be found that the overall spectral weight between 50\icm and 12\,000\icm is temperature-independent~\footnote{Since the high energy($>$~12\,000\icm) optical spectral we observed does not vary with the temperatue, we do not take the spectral weight in this area into consideration.} while the spectral weight at high frequency ($5\,000\sim12\,000\icm$) varies significantly with the temperature, suggesting that there indeed exists obvious Hund's rule coupling effect. Above $T_{CT}$, more low-frequency spectral weight transfers to high energy area ($>$~5\,000\icm) with decreasing temperature, but after the phase transition, this tendency is suppressed. The spectral weight above 5\,000\icm starts to decline, indicating the weakening of the Hund's coupling effect, which can lead to weaker correlation and smaller Fe local moment~\cite{mandal}.
%
% Figure 4
%
\begin{figure}[tb]
\includegraphics[width=1.0\columnwidth]{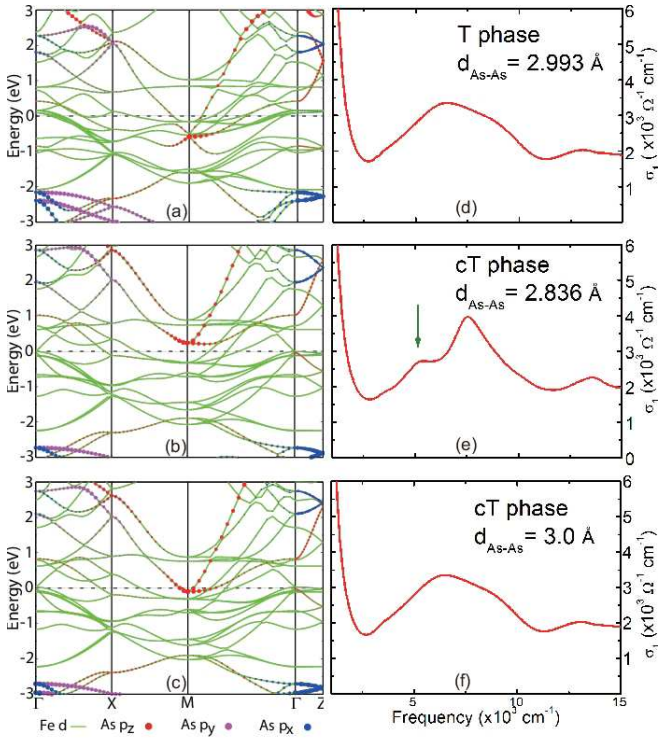}
\caption{(color on line) The calculated band structure and simulated optical conductivity in the T phase(a) (d) and cT phase (b) (e). (c) (f) are results calculated with enlarged interlayer distance and unchanged intralayer structure of that in the cT phase. See the text for more detail. }
\label{calculation}
\end{figure}

A remarkable feature we observed in the optical spectra is the newly formed mid-infrared peak around 0.4~eV after the phase transition. By comparing the optical spectral of \CCA\ which has an intrinsic As-As interlayer bond~\cite{Cheng2012}, we suppose that this new feature may be associated with the formation of As-As bond in the cT phase. To confirm this hypothesis, we have calculated the band structure and simulated the optical conductivity in different situations. The calculations are performed using density functional theory (DFT) as implemented in the Vienna $ab\ initio$ simulation package (VASP) code~\cite{Kre1996,Kre1993,Kre}. Generalized-gradient approximation (GGA) for the exchange correlation functional are used~\cite{Gene}.Throughout the work, the cutoff energy is set to be 400 eV for expanding the
wave functions into plane-wave basis£¬and the number of these k points are 16 16 8. Furthermore, the experimental parameters~\cite{diehl}($a_{T}$=3.8915~\AA, $c_{T}$=11.690~\AA; $a_{cT}$=3.9792~\AA,$c_{cT}$=10.6073~\AA) were used in the calculation. The results are shown in Fig.~\ref{calculation}, the simulated optical conductivities agree well with the observed ones in both the T and cT phase qualitatively (more detail results could be seen in Appendix ~\ref{APPENDIX C}). Then we enlarge the interlayer distance without changing the intralayer structure in the cT phase. When the distance exceeds 3.0~{\AA}, the antibonding state sinks below the Fermi level (Fig.~\ref{calculation}(c)), corresponding to the destroyed As-As bond. Simultaneously, the mid-infrared peak (as illustrated in Fig.~\ref{calculation}(e) by the green arrow) in the simulated optical conductivity disappears (Fig.~\ref{calculation}(f)). This provides strong evidence that the mid-infrared peak around 0.4eV is associated with the interlayer As-As bond.
From the optical conductivities (Fig.~\ref{optical}(b)) and their fitting results (Fig.~\ref{fit}(a)(b)), we find that, the newly formed 0.4~eV peak seems to absorb the spectral weight of the broad Drude item at low frequency and frustrate the spectral weight transfer from low to high frequency area. Hence, we infer that the As-As bond may relate to the weaker correlation, quenched magnetism as well as the disappeared bulk superconductivity in the cT phase.

According to the calculated band structure (Fig.~\ref{calculation}(a-c)), the bonding-antibonding splitting of As $p_x$/$p_y$ orbitals around the $\Gamma$ point is found to be enlarged after the cT phase transition, suggesting increased overlap between Fe $d_{xy}$ orbitals and As $p_x$/$p_y$ orbitals~\cite{Belashchenko2008} due to shorter Fe-As covalent bond induced by the formation of As-As bond~\cite{saha}. The enhanced overlap will make the electrons, especially those in $d_{xy}$ orbitals, more itinerant and greatly weakens the correlation~\cite{diehl,yin}. Since the Fe spin polarization is driven by the on-site (Hund) exchange~\cite{Belashchenko2008}, the more itinerant electrons are hardly to form the local moment. Thus, the shorter Fe-As bond ($<$~2.36~{\AA}~\cite{saha}) is responsible for the weaker correlation and quenched local moment~\cite{Mirbt2003}. On the other hand, around the M point, the As $p_z$ orbitals, which locate under the Fermi level and hybridize well with the Fe $3d$ orbitals in the T phase (Fig.~\ref{calculation}(a)), have been pushed above the Fermi level and decoupled with the Fe $3d$ orbitals after the cT phase transition (Fig.~\ref{calculation}(b)). Previous research pointed out that the As-As bond perpendicular to the Fe layer was dominated by the As $p_z$ orbitals~\cite{Hoffmann1985}, thus the empty antibonding states of As $p_z$ orbitals in Fig.~\ref{calculation}(b) correspond to the formation of As-As bond in the cT phase. Since the hybridizations between the As $p_z$ orbitals and the Fe $3d$ orbitals delivers the antiferromagnetic superexchange interaction between the nearest and next nearest Fe atoms to form the AFM order and to induce the spin fluctuation~\cite{Ma2008}, the decoupling effect caused by the newly formed interlayer As-As bond greatly frustrates the antiferromagnetic interaction and, in turn, eliminate the AFM order and spin fluctuation in the cT phase. In addition, the bulk superconductivity disappearing with the spin fluctuation supports the notion of a coupling between spin fluctuations and unconventional superconductivity in the iron pnictides~\cite{Pag,Soh2013}. Accordingly, the newly formed As-As hybridization plays a decisive role on suppressing the correlation, magnetism and superconductivity in the cT phase by affecting the Fe-As hybridiztion.

%%%%%%%%%%%%%%%%%%%%%%%%%%%%%%%%%%%%%%%%%%%%%%%%%%%%%%%%%%%%%%%%%%%%%%%%%%%%%%%
%
% Conclusions
%
In summary, we report detailed transport and optical study results on \CPFA\ single crystal, which undergoes structural phase transition from tetragonal to collapsed tetragonal phase. After the cT phase transition, the sharper reflectance edge and great spectral weight redistribution in the optical conductivity reflect suppressed spin fluctuation and weaker electron correlation. Based on the fist-principle calculation, we confirm that the newly formed feature around 0.4~eV is caused by the interlayer As-As bonding. The formation of As-As bond in the cT phase weakens the correlation, quenches the local moment and frustrates the dynamical spin fluctuation by affecting the Fe-As hybridization. Since the superconductivity in iron pnictides is widely believed to be mediated by the spin fluctuation, the degree of hybridization between iron and arsenic atoms can provide a possible explanation on the lack of bulk superconductivity in the cT phase.

%%%%%%%%%%%%%%%%%%%%%%%%%%%%%%%%%%%%%%%%%%%%%%%%%%%%%%%%%%%%%%%%%%%%%%%%%%%%%%%
%
% Acknowledgment
%

\begin{acknowledgments}
The authors thank Pengcheng Dai, Yaomin Dai, Bohong Li and Fei Cheng for useful discussion,
and we thank Yuping Sun, Minghu Fang, Hangdong Wang for assistance with X-ray diffraction measurements.
This work was supported by the NSFC (Grants No. 91121004 and 973 Projects No. 2015CB921303) and the MOST (973 Projects No. 2012CB21403, No. 2011CBA00107, No. 2012CB921302, and No. 2015CB929103).
\end{acknowledgments}

\appendix

\section{OPTICAL CONDUCTIVITY BEFORE AND AFTER THE LATTICE COLLAPSE TRANSITION}
\label{APPENDIX A}
\begin{figure}[tb]
\includegraphics[width=0.6\columnwidth]{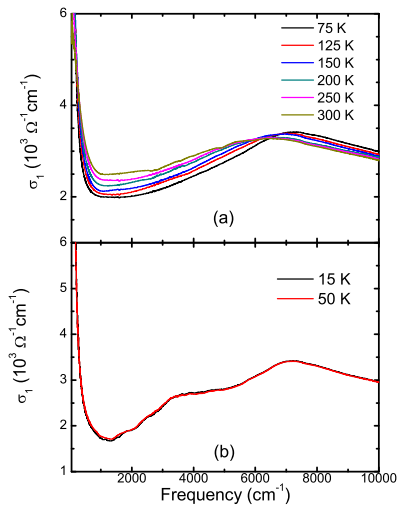}
\caption{ (color online) Optical conductivity with all measured temperature above 75~K (a) and one below 75~K (b) }
\label{Optical}
\end{figure}

In Fig.~\ref{Optical} we present all the measured optical conductivity in two pictures, one with all measured temperature above 75~K and one below 75~K. The optical conductivity of \CPFA\ in the normal state ($>$~75~K) shows typical feature of 122 iron-pnictides. The peak-like structural around 6\,000~\icm has been regarded as a high-energy pseudogap which relates to Hund's rule coupling effect. From 300~K to 75~K this peak shift from 6\,000~\icm to 7\,000~\icm. This peak shifting is common in Ba122 and Sr122 systems~\cite{WangNL}, but the range is wider in our sample. The reason for such big shifting range is still unknown, we think that this anomalous shifting may be related to the structural instability and the spin state transition in rare-earth doped \CFA~\cite{Gre2013} and needs further investigation.

\section{THE DETAIL FITTING RESULTS}
\label{APPENDIX B}
\begin{figure}[tb]
\includegraphics[width=1.0\columnwidth]{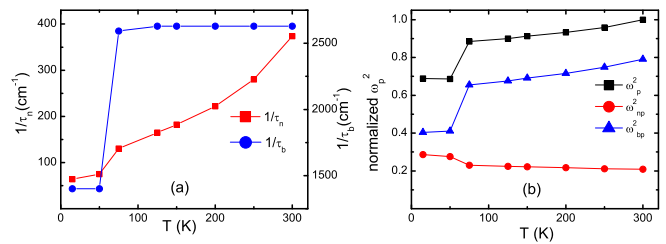}
\caption{ (color online) Temperature dependence of $\omega^{2}_{p}$ (a) normalized to the value of 300~K and scattering rate $1/\tau$ (b) of the two Drude terms of \CPFA. }
\label{Fitting}
\end{figure}

To quantitatively analysis the optical data we fitted the optical conductivity with generally used two-Drude model, the results are shown in Fig.~\ref{Fitting}. The subscript n and b stand for the narrow and broad Drude terms, respectively. From the results, we find out that, upon entering the cT phase, the plasma frequency and the scattering rate of broad Drude term shows abrupt decrease, suggesting remarkable shrinkage of broad Drude as well as the hole pockets which are caused by the formation of As-As bond.

\section{MORE CALCULATED RESULTS}
\label{APPENDIX C}
\begin{figure}[tb]
\includegraphics[width=0.6\columnwidth]{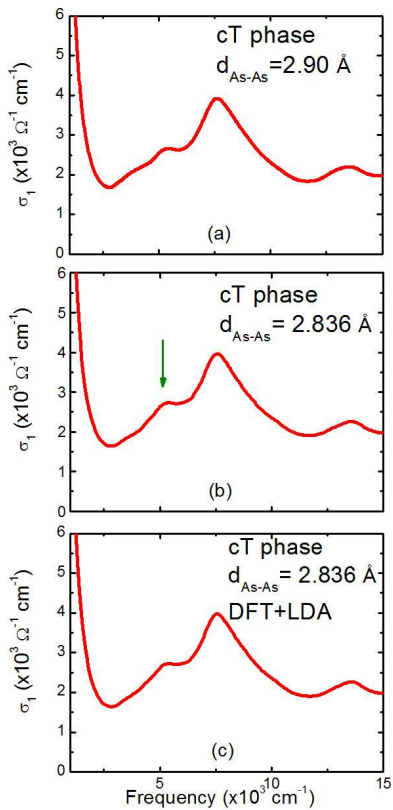}
\caption{ (color online) (a)and (b) are simulated optical conductivity with different interlayer distance. (c) is the optical conductivity calculated with local density approximation for the exchange potential.}
\label{Calculated}
\end{figure}

 To demonstrated whether the interlayer distance could greatly affect the position of new formed mid-infrared peak in the cT phase. We calculate the optical conductivity with the interlayer distance of 2.90~{\AA}, the results are shown in Fig.~\ref{Calculated}. By comparing the results of 2.83~{\AA}, we do not observe obvious shifting of the new-formed mid-infrared structure. We infer that the difference may not be caused by the different interlayer distance. Very recently, S. Mandal et al. has calculated the optical conductivity in the cT phase base on DFT+DMFT~\cite{mandal}. The calculated data precisely reproduced the experimental one. The discrepancy of our data may come from the strong correlation, which is underestimated in our calculation. However, the correlation does not affect the formation of As-As bond in the cT phase, our calculation can qualitatively reflect the nature of the sample in the cT phase.

%%%%%%%%%%%%%%%%%%%%%%%%%%%%%%%%%%%%%%%%%%%%%%%%%%%%%%%%%%%%%%%%%%%%%%%%%%%%%%%
%
% The bibliography (BibTeX)
%
%\bibliography{bib}

\begin{thebibliography}{41}%
\makeatletter
\providecommand \@ifxundefined [1]{%
 \@ifx{#1\undefined}
}%
\providecommand \@ifnum [1]{%
 \ifnum #1\expandafter \@firstoftwo
 \else \expandafter \@secondoftwo
 \fi
}%
\providecommand \@ifx [1]{%
 \ifx #1\expandafter \@firstoftwo
 \else \expandafter \@secondoftwo
 \fi
}%
\providecommand \natexlab [1]{#1}%
\providecommand \enquote  [1]{``#1''}%
\providecommand \bibnamefont  [1]{#1}%
\providecommand \bibfnamefont [1]{#1}%
\providecommand \citenamefont [1]{#1}%
\providecommand \href@noop [0]{\@secondoftwo}%
\providecommand \href [0]{\begingroup \@sanitize@url \@href}%
\providecommand \@href[1]{\@@startlink{#1}\@@href}%
\providecommand \@@href[1]{\endgroup#1\@@endlink}%
\providecommand \@sanitize@url [0]{\catcode `\\12\catcode `\$12\catcode
  `\&12\catcode `\#12\catcode `\^12\catcode `\_12\catcode `\%12\relax}%
\providecommand \@@startlink[1]{}%
\providecommand \@@endlink[0]{}%
\providecommand \url  [0]{\begingroup\@sanitize@url \@url }%
\providecommand \@url [1]{\endgroup\@href {#1}{\urlprefix }}%
\providecommand \urlprefix  [0]{URL }%
\providecommand \Eprint [0]{\href }%
\providecommand \doibase [0]{http://dx.doi.org/}%
\providecommand \selectlanguage [0]{\@gobble}%
\providecommand \bibinfo  [0]{\@secondoftwo}%
\providecommand \bibfield  [0]{\@secondoftwo}%
\providecommand \translation [1]{[#1]}%
\providecommand \BibitemOpen [0]{}%
\providecommand \bibitemStop [0]{}%
\providecommand \bibitemNoStop [0]{.\EOS\space}%
\providecommand \EOS [0]{\spacefactor3000\relax}%
\providecommand \BibitemShut  [1]{\csname bibitem#1\endcsname}%
\let\auto@bib@innerbib\@empty
%</preamble>
\bibitem [{\citenamefont {Lee}\ \emph {et~al.}(2006)\citenamefont {Lee},
  \citenamefont {Nagaosa},\ and\ \citenamefont {Wen}}]{Lee}%
  \BibitemOpen
  \bibfield  {author} {\bibinfo {author} {\bibfnamefont {P.~A.}\ \bibnamefont
  {Lee}}, \bibinfo {author} {\bibfnamefont {N.}~\bibnamefont {Nagaosa}}, \ and\
  \bibinfo {author} {\bibfnamefont {X.-G.}\ \bibnamefont {Wen}},\ }\href
  {\doibase 10.1103/RevModPhys.78.17} {\bibfield  {journal} {\bibinfo
  {journal} {Rev. Mod. Phys.}\ }\textbf {\bibinfo {volume} {78}},\ \bibinfo
  {pages} {17} (\bibinfo {year} {2006})}\BibitemShut {NoStop}%
\bibitem [{\citenamefont {Johnston}(2010)}]{Johnston2010}%
  \BibitemOpen
  \bibfield  {author} {\bibinfo {author} {\bibfnamefont {D.~C.}\ \bibnamefont
  {Johnston}},\ }\href {\doibase 10.1080/00018732.2010.513480} {\bibfield
  {journal} {\bibinfo  {journal} {Advances in Physics}\ }\textbf {\bibinfo
  {volume} {59}},\ \bibinfo {pages} {803} (\bibinfo {year} {2010})}\BibitemShut
  {NoStop}%
\bibitem [{\citenamefont {Paglione}\ and\ \citenamefont {Greene}(2010)}]{Pag}%
  \BibitemOpen
  \bibfield  {author} {\bibinfo {author} {\bibfnamefont {J.}~\bibnamefont
  {Paglione}}\ and\ \bibinfo {author} {\bibfnamefont {R.~L.}\ \bibnamefont
  {Greene}},\ }\href {\doibase 10.1038/nphys1759} {\bibfield  {journal}
  {\bibinfo  {journal} {Nat. Phys.}\ }\textbf {\bibinfo {volume} {6}},\
  \bibinfo {pages} {645} (\bibinfo {year} {2010})}\BibitemShut {NoStop}%
\bibitem [{\citenamefont {Belashchenko}\ and\ \citenamefont
  {Antropov}(2008)}]{Belashchenko2008}%
  \BibitemOpen
  \bibfield  {author} {\bibinfo {author} {\bibfnamefont {K.~D.}\ \bibnamefont
  {Belashchenko}}\ and\ \bibinfo {author} {\bibfnamefont {V.~P.}\ \bibnamefont
  {Antropov}},\ }\href {\doibase 10.1103/PhysRevB.78.212505} {\bibfield
  {journal} {\bibinfo  {journal} {Phys. Rev. B}\ }\textbf {\bibinfo {volume}
  {78}},\ \bibinfo {pages} {212505} (\bibinfo {year} {2008})}\BibitemShut
  {NoStop}%
\bibitem [{\citenamefont {Diehl}\ \emph {et~al.}(2014)\citenamefont {Diehl},
  \citenamefont {Backes}, \citenamefont {Guterding}, \citenamefont {Jeschke},\
  and\ \citenamefont {Valent\'{\i}}}]{diehl}%
  \BibitemOpen
  \bibfield  {author} {\bibinfo {author} {\bibfnamefont {J.}~\bibnamefont
  {Diehl}}, \bibinfo {author} {\bibfnamefont {S.}~\bibnamefont {Backes}},
  \bibinfo {author} {\bibfnamefont {D.}~\bibnamefont {Guterding}}, \bibinfo
  {author} {\bibfnamefont {H.~O.}\ \bibnamefont {Jeschke}}, \ and\ \bibinfo
  {author} {\bibfnamefont {R.}~\bibnamefont {Valent\'{\i}}},\ }\href {\doibase
  10.1103/PhysRevB.90.085110} {\bibfield  {journal} {\bibinfo  {journal} {Phys.
  Rev. B}\ }\textbf {\bibinfo {volume} {90}},\ \bibinfo {pages} {085110}
  (\bibinfo {year} {2014})}\BibitemShut {NoStop}%
\bibitem [{\citenamefont {Ma}\ \emph {et~al.}(2008)\citenamefont {Ma},
  \citenamefont {Lu},\ and\ \citenamefont {Xiang}}]{Ma2008}%
  \BibitemOpen
  \bibfield  {author} {\bibinfo {author} {\bibfnamefont {F.}~\bibnamefont
  {Ma}}, \bibinfo {author} {\bibfnamefont {Z.-Y.}\ \bibnamefont {Lu}}, \ and\
  \bibinfo {author} {\bibfnamefont {T.}~\bibnamefont {Xiang}},\ }\href
  {\doibase 10.1103/PhysRevB.78.224517} {\bibfield  {journal} {\bibinfo
  {journal} {Phys. Rev. B}\ }\textbf {\bibinfo {volume} {78}},\ \bibinfo
  {pages} {224517} (\bibinfo {year} {2008})}\BibitemShut {NoStop}%
\bibitem [{\citenamefont {Mirbt}\ \emph {et~al.}(2003)\citenamefont {Mirbt},
  \citenamefont {Sanyal}, \citenamefont {Isheden},\ and\ \citenamefont
  {Johansson}}]{Mirbt2003}%
  \BibitemOpen
  \bibfield  {author} {\bibinfo {author} {\bibfnamefont {S.}~\bibnamefont
  {Mirbt}}, \bibinfo {author} {\bibfnamefont {B.}~\bibnamefont {Sanyal}},
  \bibinfo {author} {\bibfnamefont {C.}~\bibnamefont {Isheden}}, \ and\
  \bibinfo {author} {\bibfnamefont {B.}~\bibnamefont {Johansson}},\ }\href
  {\doibase 10.1103/PhysRevB.67.155421} {\bibfield  {journal} {\bibinfo
  {journal} {Phys. Rev. B}\ }\textbf {\bibinfo {volume} {67}},\ \bibinfo
  {pages} {155421} (\bibinfo {year} {2003})}\BibitemShut {NoStop}%
\bibitem [{\citenamefont {Saparov}\ \emph {et~al.}(2014)\citenamefont
  {Saparov}, \citenamefont {Cantoni}, \citenamefont {Pan}, \citenamefont
  {Hogan}, \citenamefont {Ratcliff}, \citenamefont {Wilson}, \citenamefont
  {Fritsch}, \citenamefont {Gaulin},\ and\ \citenamefont {Sefat}}]{Sapa2014}%
  \BibitemOpen
  \bibfield  {author} {\bibinfo {author} {\bibfnamefont {B.}~\bibnamefont
  {Saparov}}, \bibinfo {author} {\bibfnamefont {C.}~\bibnamefont {Cantoni}},
  \bibinfo {author} {\bibfnamefont {M.}~\bibnamefont {Pan}}, \bibinfo {author}
  {\bibfnamefont {T.~C.}\ \bibnamefont {Hogan}}, \bibinfo {author}
  {\bibfnamefont {W.}~\bibnamefont {Ratcliff}}, \bibinfo {author}
  {\bibfnamefont {S.~D.}\ \bibnamefont {Wilson}}, \bibinfo {author}
  {\bibfnamefont {K.}~\bibnamefont {Fritsch}}, \bibinfo {author} {\bibfnamefont
  {B.~D.}\ \bibnamefont {Gaulin}}, \ and\ \bibinfo {author} {\bibfnamefont
  {A.~S.}\ \bibnamefont {Sefat}},\ }\href {\doibase 10.1038/srep04120}
  {\bibfield  {journal} {\bibinfo  {journal} {Sci. rep.}\ }\textbf {\bibinfo
  {volume} {4}},\ \bibinfo {pages} {4120} (\bibinfo {year} {2014})}\BibitemShut
  {NoStop}%
\bibitem [{\citenamefont {Tompsett}\ and\ \citenamefont
  {Lonzarich}(2009)}]{Tom}%
  \BibitemOpen
  \bibfield  {author} {\bibinfo {author} {\bibfnamefont {D.~A.}\ \bibnamefont
  {Tompsett}}\ and\ \bibinfo {author} {\bibfnamefont {G.~G.}\ \bibnamefont
  {Lonzarich}},\ }\href@noop {} {\bibfield  {journal} {\bibinfo  {journal}
  {arXiv:0902.4859}\ } (\bibinfo {year} {2009})}\BibitemShut {NoStop}%
\bibitem [{\citenamefont {Gretarsson}\ \emph {et~al.}(2013)\citenamefont
  {Gretarsson}, \citenamefont {Saha}, \citenamefont {Drye}, \citenamefont
  {Paglione}, \citenamefont {Kim}, \citenamefont {Casa}, \citenamefont {Gog},
  \citenamefont {Wu}, \citenamefont {Julian},\ and\ \citenamefont
  {Kim}}]{Gre2013}%
  \BibitemOpen
  \bibfield  {author} {\bibinfo {author} {\bibfnamefont {H.}~\bibnamefont
  {Gretarsson}}, \bibinfo {author} {\bibfnamefont {S.~R.}\ \bibnamefont
  {Saha}}, \bibinfo {author} {\bibfnamefont {T.}~\bibnamefont {Drye}}, \bibinfo
  {author} {\bibfnamefont {J.}~\bibnamefont {Paglione}}, \bibinfo {author}
  {\bibfnamefont {J.}~\bibnamefont {Kim}}, \bibinfo {author} {\bibfnamefont
  {D.}~\bibnamefont {Casa}}, \bibinfo {author} {\bibfnamefont {T.}~\bibnamefont
  {Gog}}, \bibinfo {author} {\bibfnamefont {W.}~\bibnamefont {Wu}}, \bibinfo
  {author} {\bibfnamefont {S.~R.}\ \bibnamefont {Julian}}, \ and\ \bibinfo
  {author} {\bibfnamefont {Y.-J.}\ \bibnamefont {Kim}},\ }\href {\doibase
  10.1103/PhysRevLett.110.047003} {\bibfield  {journal} {\bibinfo  {journal}
  {Phys. Rev. Lett.}\ }\textbf {\bibinfo {volume} {110}},\ \bibinfo {pages}
  {047003} (\bibinfo {year} {2013})}\BibitemShut {NoStop}%
\bibitem [{\citenamefont {Canfield}\ \emph {et~al.}(2009)\citenamefont
  {Canfield}, \citenamefont {Bud¡¯ko}, \citenamefont {Ni}, \citenamefont
  {Kreyssig}, \citenamefont {Goldman}, \citenamefont {McQueeney}, \citenamefont
  {Torikachvili}, \citenamefont {Argyriou}, \citenamefont {Luke},\ and\
  \citenamefont {Yu}}]{can}%
  \BibitemOpen
  \bibfield  {author} {\bibinfo {author} {\bibfnamefont {P.}~\bibnamefont
  {Canfield}}, \bibinfo {author} {\bibfnamefont {S.}~\bibnamefont {Bud¡¯ko}},
  \bibinfo {author} {\bibfnamefont {N.}~\bibnamefont {Ni}}, \bibinfo {author}
  {\bibfnamefont {A.}~\bibnamefont {Kreyssig}}, \bibinfo {author}
  {\bibfnamefont {A.}~\bibnamefont {Goldman}}, \bibinfo {author} {\bibfnamefont
  {R.}~\bibnamefont {McQueeney}}, \bibinfo {author} {\bibfnamefont
  {M.}~\bibnamefont {Torikachvili}}, \bibinfo {author} {\bibfnamefont
  {D.}~\bibnamefont {Argyriou}}, \bibinfo {author} {\bibfnamefont
  {G.}~\bibnamefont {Luke}}, \ and\ \bibinfo {author} {\bibfnamefont
  {W.}~\bibnamefont {Yu}},\ }\href {\doibase 10.1016/j.physc.2009.03.033}
  {\bibfield  {journal} {\bibinfo  {journal} {Physica C: Superconductivity}\
  }\textbf {\bibinfo {volume} {469}},\ \bibinfo {pages} {404} (\bibinfo {year}
  {2009})}\BibitemShut {NoStop}%
\bibitem [{\citenamefont {Torikachvili}\ \emph {et~al.}(2008)\citenamefont
  {Torikachvili}, \citenamefont {Bud'ko}, \citenamefont {Ni},\ and\
  \citenamefont {Canfield}}]{Tori2008}%
  \BibitemOpen
  \bibfield  {author} {\bibinfo {author} {\bibfnamefont {M.~S.}\ \bibnamefont
  {Torikachvili}}, \bibinfo {author} {\bibfnamefont {S.~L.}\ \bibnamefont
  {Bud'ko}}, \bibinfo {author} {\bibfnamefont {N.}~\bibnamefont {Ni}}, \ and\
  \bibinfo {author} {\bibfnamefont {P.~C.}\ \bibnamefont {Canfield}},\ }\href
  {\doibase 10.1103/PhysRevLett.101.057006} {\bibfield  {journal} {\bibinfo
  {journal} {Phys. Rev. Lett.}\ }\textbf {\bibinfo {volume} {101}},\ \bibinfo
  {pages} {057006} (\bibinfo {year} {2008})}\BibitemShut {NoStop}%
\bibitem [{\citenamefont {Yu}\ \emph {et~al.}(2009)\citenamefont {Yu},
  \citenamefont {Aczel}, \citenamefont {Williams}, \citenamefont {Bud'ko},
  \citenamefont {Ni}, \citenamefont {Canfield},\ and\ \citenamefont
  {Luke}}]{Yu}%
  \BibitemOpen
  \bibfield  {author} {\bibinfo {author} {\bibfnamefont {W.}~\bibnamefont
  {Yu}}, \bibinfo {author} {\bibfnamefont {A.~A.}\ \bibnamefont {Aczel}},
  \bibinfo {author} {\bibfnamefont {T.~J.}\ \bibnamefont {Williams}}, \bibinfo
  {author} {\bibfnamefont {S.~L.}\ \bibnamefont {Bud'ko}}, \bibinfo {author}
  {\bibfnamefont {N.}~\bibnamefont {Ni}}, \bibinfo {author} {\bibfnamefont
  {P.~C.}\ \bibnamefont {Canfield}}, \ and\ \bibinfo {author} {\bibfnamefont
  {G.~M.}\ \bibnamefont {Luke}},\ }\href {\doibase 10.1103/PhysRevB.79.020511}
  {\bibfield  {journal} {\bibinfo  {journal} {Phys. Rev. B}\ }\textbf {\bibinfo
  {volume} {79}},\ \bibinfo {pages} {020511} (\bibinfo {year}
  {2009})}\BibitemShut {NoStop}%
\bibitem [{\citenamefont {Saha}\ \emph {et~al.}(2012)\citenamefont {Saha},
  \citenamefont {Butch}, \citenamefont {Drye}, \citenamefont {Magill},
  \citenamefont {Ziemak}, \citenamefont {Kirshenbaum}, \citenamefont {Zavalij},
  \citenamefont {Lynn},\ and\ \citenamefont {Paglione}}]{saha}%
  \BibitemOpen
  \bibfield  {author} {\bibinfo {author} {\bibfnamefont {S.~R.}\ \bibnamefont
  {Saha}}, \bibinfo {author} {\bibfnamefont {N.~P.}\ \bibnamefont {Butch}},
  \bibinfo {author} {\bibfnamefont {T.}~\bibnamefont {Drye}}, \bibinfo {author}
  {\bibfnamefont {J.}~\bibnamefont {Magill}}, \bibinfo {author} {\bibfnamefont
  {S.}~\bibnamefont {Ziemak}}, \bibinfo {author} {\bibfnamefont
  {K.}~\bibnamefont {Kirshenbaum}}, \bibinfo {author} {\bibfnamefont {P.~Y.}\
  \bibnamefont {Zavalij}}, \bibinfo {author} {\bibfnamefont {J.~W.}\
  \bibnamefont {Lynn}}, \ and\ \bibinfo {author} {\bibfnamefont
  {J.}~\bibnamefont {Paglione}},\ }\href {\doibase 10.1103/PhysRevB.85.024525}
  {\bibfield  {journal} {\bibinfo  {journal} {Phys. Rev. B}\ }\textbf {\bibinfo
  {volume} {85}},\ \bibinfo {pages} {024525} (\bibinfo {year}
  {2012})}\BibitemShut {NoStop}%
\bibitem [{\citenamefont {Ma}\ \emph {et~al.}(2013)\citenamefont {Ma},
  \citenamefont {Ji}, \citenamefont {Dai}, \citenamefont {Saha}, \citenamefont
  {Drye}, \citenamefont {Paglione},\ and\ \citenamefont {Yu}}]{Ma2013}%
  \BibitemOpen
  \bibfield  {author} {\bibinfo {author} {\bibfnamefont {L.}~\bibnamefont
  {Ma}}, \bibinfo {author} {\bibfnamefont {G.-F.}\ \bibnamefont {Ji}}, \bibinfo
  {author} {\bibfnamefont {J.}~\bibnamefont {Dai}}, \bibinfo {author}
  {\bibfnamefont {S.~R.}\ \bibnamefont {Saha}}, \bibinfo {author}
  {\bibfnamefont {T.}~\bibnamefont {Drye}}, \bibinfo {author} {\bibfnamefont
  {J.}~\bibnamefont {Paglione}}, \ and\ \bibinfo {author} {\bibfnamefont
  {W.-Q.}\ \bibnamefont {Yu}},\ }\href {\doibase 10.1088/1674-1056/22/5/057401}
  {\bibfield  {journal} {\bibinfo  {journal} {Chin. Phys. B}\ }\textbf
  {\bibinfo {volume} {22}},\ \bibinfo {pages} {057401} (\bibinfo {year}
  {2013})}\BibitemShut {NoStop}%
\bibitem [{\citenamefont {Soh}\ \emph {et~al.}(2013)\citenamefont {Soh},
  \citenamefont {Tucker}, \citenamefont {Pratt}, \citenamefont {Abernathy},
  \citenamefont {Stone}, \citenamefont {Ran}, \citenamefont {Bud'ko},
  \citenamefont {Canfield}, \citenamefont {Kreyssig}, \citenamefont
  {McQueeney},\ and\ \citenamefont {Goldman}}]{Soh2013}%
  \BibitemOpen
  \bibfield  {author} {\bibinfo {author} {\bibfnamefont {J.~H.}\ \bibnamefont
  {Soh}}, \bibinfo {author} {\bibfnamefont {G.~S.}\ \bibnamefont {Tucker}},
  \bibinfo {author} {\bibfnamefont {D.~K.}\ \bibnamefont {Pratt}}, \bibinfo
  {author} {\bibfnamefont {D.~L.}\ \bibnamefont {Abernathy}}, \bibinfo {author}
  {\bibfnamefont {M.~B.}\ \bibnamefont {Stone}}, \bibinfo {author}
  {\bibfnamefont {S.}~\bibnamefont {Ran}}, \bibinfo {author} {\bibfnamefont
  {S.~L.}\ \bibnamefont {Bud'ko}}, \bibinfo {author} {\bibfnamefont {P.~C.}\
  \bibnamefont {Canfield}}, \bibinfo {author} {\bibfnamefont {A.}~\bibnamefont
  {Kreyssig}}, \bibinfo {author} {\bibfnamefont {R.~J.}\ \bibnamefont
  {McQueeney}}, \ and\ \bibinfo {author} {\bibfnamefont {A.~I.}\ \bibnamefont
  {Goldman}},\ }\href {\doibase 10.1103/PhysRevLett.111.227002} {\bibfield
  {journal} {\bibinfo  {journal} {Phys. Rev. Lett.}\ }\textbf {\bibinfo
  {volume} {111}},\ \bibinfo {pages} {227002} (\bibinfo {year}
  {2013})}\BibitemShut {NoStop}%
\bibitem [{\citenamefont {Kasahara}\ \emph {et~al.}(2011)\citenamefont
  {Kasahara}, \citenamefont {Shibauchi}, \citenamefont {Hashimoto},
  \citenamefont {Nakai}, \citenamefont {Ikeda}, \citenamefont {Terashima},\
  and\ \citenamefont {Matsuda}}]{kas}%
  \BibitemOpen
  \bibfield  {author} {\bibinfo {author} {\bibfnamefont {S.}~\bibnamefont
  {Kasahara}}, \bibinfo {author} {\bibfnamefont {T.}~\bibnamefont {Shibauchi}},
  \bibinfo {author} {\bibfnamefont {K.}~\bibnamefont {Hashimoto}}, \bibinfo
  {author} {\bibfnamefont {Y.}~\bibnamefont {Nakai}}, \bibinfo {author}
  {\bibfnamefont {H.}~\bibnamefont {Ikeda}}, \bibinfo {author} {\bibfnamefont
  {T.}~\bibnamefont {Terashima}}, \ and\ \bibinfo {author} {\bibfnamefont
  {Y.}~\bibnamefont {Matsuda}},\ }\href {\doibase 10.1103/PhysRevB.83.060505}
  {\bibfield  {journal} {\bibinfo  {journal} {Phys. Rev. B}\ }\textbf {\bibinfo
  {volume} {83}},\ \bibinfo {pages} {060505} (\bibinfo {year}
  {2011})}\BibitemShut {NoStop}%
\bibitem [{\citenamefont {Yildirim}(2009)}]{Yildirim2009}%
  \BibitemOpen
  \bibfield  {author} {\bibinfo {author} {\bibfnamefont {T.}~\bibnamefont
  {Yildirim}},\ }\href {\doibase 10.1103/PhysRevLett.102.037003} {\bibfield
  {journal} {\bibinfo  {journal} {Phys. Rev. Lett.}\ }\textbf {\bibinfo
  {volume} {102}},\ \bibinfo {pages} {037003} (\bibinfo {year}
  {2009})}\BibitemShut {NoStop}%
\bibitem [{\citenamefont {Wang}\ \emph {et~al.}(2014)\citenamefont {Wang},
  \citenamefont {Wang}, \citenamefont {Dong}, \citenamefont {Chen},\ and\
  \citenamefont {Wang}}]{wang}%
  \BibitemOpen
  \bibfield  {author} {\bibinfo {author} {\bibfnamefont {X.~B.}\ \bibnamefont
  {Wang}}, \bibinfo {author} {\bibfnamefont {H.~P.}\ \bibnamefont {Wang}},
  \bibinfo {author} {\bibfnamefont {T.}~\bibnamefont {Dong}}, \bibinfo {author}
  {\bibfnamefont {R.~Y.}\ \bibnamefont {Chen}}, \ and\ \bibinfo {author}
  {\bibfnamefont {N.~L.}\ \bibnamefont {Wang}},\ }\href {\doibase
  10.1103/PhysRevB.90.144513} {\bibfield  {journal} {\bibinfo  {journal} {Phys.
  Rev. B}\ }\textbf {\bibinfo {volume} {90}},\ \bibinfo {pages} {144513}
  (\bibinfo {year} {2014})}\BibitemShut {NoStop}%
\bibitem [{\citenamefont {Ronning}\ \emph {et~al.}(2008)\citenamefont
  {Ronning}, \citenamefont {Klimczuk}, \citenamefont {Bauer}, \citenamefont
  {Volz},\ and\ \citenamefont {Thompson}}]{ron}%
  \BibitemOpen
  \bibfield  {author} {\bibinfo {author} {\bibfnamefont {F.}~\bibnamefont
  {Ronning}}, \bibinfo {author} {\bibfnamefont {T.}~\bibnamefont {Klimczuk}},
  \bibinfo {author} {\bibfnamefont {E.~D.}\ \bibnamefont {Bauer}}, \bibinfo
  {author} {\bibfnamefont {H.}~\bibnamefont {Volz}}, \ and\ \bibinfo {author}
  {\bibfnamefont {J.~D.}\ \bibnamefont {Thompson}},\ }\href {\doibase
  10.1088/0953-8984/20/32/322201} {\bibfield  {journal} {\bibinfo  {journal}
  {J. Phys. Condens. Matter}\ }\textbf {\bibinfo {volume} {20}},\ \bibinfo
  {pages} {322201} (\bibinfo {year} {2008})}\BibitemShut {NoStop}%
\bibitem [{\citenamefont {Homes}\ \emph {et~al.}(1993)\citenamefont {Homes},
  \citenamefont {Reedyk}, \citenamefont {Cradles},\ and\ \citenamefont
  {Timusk}}]{Chri}%
  \BibitemOpen
  \bibfield  {author} {\bibinfo {author} {\bibfnamefont {C.~C.}\ \bibnamefont
  {Homes}}, \bibinfo {author} {\bibfnamefont {M.}~\bibnamefont {Reedyk}},
  \bibinfo {author} {\bibfnamefont {D.~A.}\ \bibnamefont {Cradles}}, \ and\
  \bibinfo {author} {\bibfnamefont {T.}~\bibnamefont {Timusk}},\ }\href
  {\doibase 10.1364/AO.32.002976} {\bibfield  {journal} {\bibinfo  {journal}
  {Appl. Opt.}\ }\textbf {\bibinfo {volume} {32}},\ \bibinfo {pages} {2976}
  (\bibinfo {year} {1993})}\BibitemShut {NoStop}%
\bibitem [{\citenamefont {Dai}\ \emph {et~al.}(2014)\citenamefont {Dai},
  \citenamefont {Akrap}, \citenamefont {Schneeloch}, \citenamefont {Zhong},
  \citenamefont {Liu}, \citenamefont {Gu}, \citenamefont {Li},\ and\
  \citenamefont {Homes}}]{Dai}%
  \BibitemOpen
  \bibfield  {author} {\bibinfo {author} {\bibfnamefont {Y.~M.}\ \bibnamefont
  {Dai}}, \bibinfo {author} {\bibfnamefont {A.}~\bibnamefont {Akrap}}, \bibinfo
  {author} {\bibfnamefont {J.}~\bibnamefont {Schneeloch}}, \bibinfo {author}
  {\bibfnamefont {R.~D.}\ \bibnamefont {Zhong}}, \bibinfo {author}
  {\bibfnamefont {T.~S.}\ \bibnamefont {Liu}}, \bibinfo {author} {\bibfnamefont
  {G.~D.}\ \bibnamefont {Gu}}, \bibinfo {author} {\bibfnamefont
  {Q.}~\bibnamefont {Li}}, \ and\ \bibinfo {author} {\bibfnamefont {C.~C.}\
  \bibnamefont {Homes}},\ }\href {\doibase 10.1103/PhysRevB.90.121114}
  {\bibfield  {journal} {\bibinfo  {journal} {Phys. Rev. B}\ }\textbf {\bibinfo
  {volume} {90}},\ \bibinfo {pages} {121114} (\bibinfo {year}
  {2014})}\BibitemShut {NoStop}%
\bibitem [{\citenamefont {Lv}\ \emph {et~al.}(2011)\citenamefont {Lv},
  \citenamefont {Deng}, \citenamefont {Gooch}, \citenamefont {Wei},
  \citenamefont {Sun}, \citenamefont {Meen}, \citenamefont {Xue}, \citenamefont
  {Lorenz},\ and\ \citenamefont {Chu}}]{Lv2011}%
  \BibitemOpen
  \bibfield  {author} {\bibinfo {author} {\bibfnamefont {B.}~\bibnamefont
  {Lv}}, \bibinfo {author} {\bibfnamefont {L.}~\bibnamefont {Deng}}, \bibinfo
  {author} {\bibfnamefont {M.}~\bibnamefont {Gooch}}, \bibinfo {author}
  {\bibfnamefont {F.}~\bibnamefont {Wei}}, \bibinfo {author} {\bibfnamefont
  {Y.}~\bibnamefont {Sun}}, \bibinfo {author} {\bibfnamefont {J.~K.}\
  \bibnamefont {Meen}}, \bibinfo {author} {\bibfnamefont {Y.-Y.}\ \bibnamefont
  {Xue}}, \bibinfo {author} {\bibfnamefont {B.}~\bibnamefont {Lorenz}}, \ and\
  \bibinfo {author} {\bibfnamefont {C.-W.}\ \bibnamefont {Chu}},\ }\href
  {\doibase 10.1073/pnas.1112150108} {\bibfield  {journal} {\bibinfo  {journal}
  {Proc. Natl. Acad. Sci. U. S. A.}\ }\textbf {\bibinfo {volume} {108}},\
  \bibinfo {pages} {15705} (\bibinfo {year} {2011})}\BibitemShut {NoStop}%
\bibitem [{\citenamefont {Gofryk}\ \emph {et~al.}(2014)\citenamefont {Gofryk},
  \citenamefont {Pan}, \citenamefont {Cantoni}, \citenamefont {Saparov},
  \citenamefont {Mitchell},\ and\ \citenamefont {Sefat}}]{Gofry2014}%
  \BibitemOpen
  \bibfield  {author} {\bibinfo {author} {\bibfnamefont {K.}~\bibnamefont
  {Gofryk}}, \bibinfo {author} {\bibfnamefont {M.}~\bibnamefont {Pan}},
  \bibinfo {author} {\bibfnamefont {C.}~\bibnamefont {Cantoni}}, \bibinfo
  {author} {\bibfnamefont {B.}~\bibnamefont {Saparov}}, \bibinfo {author}
  {\bibfnamefont {J.~E.}\ \bibnamefont {Mitchell}}, \ and\ \bibinfo {author}
  {\bibfnamefont {A.~S.}\ \bibnamefont {Sefat}},\ }\href {\doibase
  10.1103/PhysRevLett.112.047005} {\bibfield  {journal} {\bibinfo  {journal}
  {Phys. Rev. Lett.}\ }\textbf {\bibinfo {volume} {112}},\ \bibinfo {pages}
  {047005} (\bibinfo {year} {2014})}\BibitemShut {NoStop}%
\bibitem [{\citenamefont {Basov}\ and\ \citenamefont
  {Timusk}(2005)}]{Basov2005}%
  \BibitemOpen
  \bibfield  {author} {\bibinfo {author} {\bibfnamefont {D.}~\bibnamefont
  {Basov}}\ and\ \bibinfo {author} {\bibfnamefont {T.}~\bibnamefont {Timusk}},\
  }\href {\doibase 10.1103/RevModPhys.77.721} {\bibfield  {journal} {\bibinfo
  {journal} {Rev. Mod. Phys.}\ }\textbf {\bibinfo {volume} {77}},\ \bibinfo
  {pages} {721} (\bibinfo {year} {2005})}\BibitemShut {NoStop}%
\bibitem [{\citenamefont {Pratt}\ \emph {et~al.}(2009)\citenamefont {Pratt},
  \citenamefont {Zhao}, \citenamefont {Kimber}, \citenamefont {Hiess},
  \citenamefont {Argyriou}, \citenamefont {Broholm}, \citenamefont {Kreyssig},
  \citenamefont {Nandi}, \citenamefont {Bud'ko}, \citenamefont {Ni},
  \citenamefont {Canfield}, \citenamefont {McQueeney},\ and\ \citenamefont
  {Goldman}}]{Pratt}%
  \BibitemOpen
  \bibfield  {author} {\bibinfo {author} {\bibfnamefont {D.~K.}\ \bibnamefont
  {Pratt}}, \bibinfo {author} {\bibfnamefont {Y.}~\bibnamefont {Zhao}},
  \bibinfo {author} {\bibfnamefont {S.~A.~J.}\ \bibnamefont {Kimber}}, \bibinfo
  {author} {\bibfnamefont {A.}~\bibnamefont {Hiess}}, \bibinfo {author}
  {\bibfnamefont {D.~N.}\ \bibnamefont {Argyriou}}, \bibinfo {author}
  {\bibfnamefont {C.}~\bibnamefont {Broholm}}, \bibinfo {author} {\bibfnamefont
  {A.}~\bibnamefont {Kreyssig}}, \bibinfo {author} {\bibfnamefont
  {S.}~\bibnamefont {Nandi}}, \bibinfo {author} {\bibfnamefont {S.~L.}\
  \bibnamefont {Bud'ko}}, \bibinfo {author} {\bibfnamefont {N.}~\bibnamefont
  {Ni}}, \bibinfo {author} {\bibfnamefont {P.~C.}\ \bibnamefont {Canfield}},
  \bibinfo {author} {\bibfnamefont {R.~J.}\ \bibnamefont {McQueeney}}, \ and\
  \bibinfo {author} {\bibfnamefont {A.~I.}\ \bibnamefont {Goldman}},\ }\href
  {\doibase 10.1103/PhysRevB.79.060510} {\bibfield  {journal} {\bibinfo
  {journal} {Phys. Rev. B}\ }\textbf {\bibinfo {volume} {79}},\ \bibinfo
  {pages} {060510} (\bibinfo {year} {2009})}\BibitemShut {NoStop}%
\bibitem [{\citenamefont {Wu}\ \emph {et~al.}(2008)\citenamefont {Wu},
  \citenamefont {Khalifah}, \citenamefont {Mandrus},\ and\ \citenamefont
  {Wang}}]{Wu}%
  \BibitemOpen
  \bibfield  {author} {\bibinfo {author} {\bibfnamefont {D.}~\bibnamefont
  {Wu}}, \bibinfo {author} {\bibfnamefont {P.~G.}\ \bibnamefont {Khalifah}},
  \bibinfo {author} {\bibfnamefont {D.~G.}\ \bibnamefont {Mandrus}}, \ and\
  \bibinfo {author} {\bibfnamefont {N.~L.}\ \bibnamefont {Wang}},\ }\href
  {\doibase 10.1088/0953-8984/20/32/325204} {\bibfield  {journal} {\bibinfo
  {journal} {J. Phys. Condens. Matter}\ }\textbf {\bibinfo {volume} {20}},\
  \bibinfo {pages} {325204} (\bibinfo {year} {2008})}\BibitemShut {NoStop}%
\bibitem [{\citenamefont {Marsik}\ \emph {et~al.}(2013)\citenamefont {Marsik},
  \citenamefont {Wang}, \citenamefont {R\"ossle}, \citenamefont {Yazdi-Rizi},
  \citenamefont {Schuster}, \citenamefont {Kim}, \citenamefont {Dubroka},
  \citenamefont {Munzar}, \citenamefont {Wolf}, \citenamefont {Chen},\ and\
  \citenamefont {Bernhard}}]{marsik}%
  \BibitemOpen
  \bibfield  {author} {\bibinfo {author} {\bibfnamefont {P.}~\bibnamefont
  {Marsik}}, \bibinfo {author} {\bibfnamefont {C.~N.}\ \bibnamefont {Wang}},
  \bibinfo {author} {\bibfnamefont {M.}~\bibnamefont {R\"ossle}}, \bibinfo
  {author} {\bibfnamefont {M.}~\bibnamefont {Yazdi-Rizi}}, \bibinfo {author}
  {\bibfnamefont {R.}~\bibnamefont {Schuster}}, \bibinfo {author}
  {\bibfnamefont {K.~W.}\ \bibnamefont {Kim}}, \bibinfo {author} {\bibfnamefont
  {A.}~\bibnamefont {Dubroka}}, \bibinfo {author} {\bibfnamefont
  {D.}~\bibnamefont {Munzar}}, \bibinfo {author} {\bibfnamefont
  {T.}~\bibnamefont {Wolf}}, \bibinfo {author} {\bibfnamefont {X.~H.}\
  \bibnamefont {Chen}}, \ and\ \bibinfo {author} {\bibfnamefont
  {C.}~\bibnamefont {Bernhard}},\ }\href {\doibase 10.1103/PhysRevB.88.180508}
  {\bibfield  {journal} {\bibinfo  {journal} {Phys. Rev. B}\ }\textbf {\bibinfo
  {volume} {88}},\ \bibinfo {pages} {180508} (\bibinfo {year}
  {2013})}\BibitemShut {NoStop}%
\bibitem [{\citenamefont {Calder\'on}\ \emph {et~al.}(2014)\citenamefont
  {Calder\'on}, \citenamefont {Medici}, \citenamefont {Valenzuela},\ and\
  \citenamefont {Bascones}}]{cal}%
  \BibitemOpen
  \bibfield  {author} {\bibinfo {author} {\bibfnamefont {M.~J.}\ \bibnamefont
  {Calder\'on}}, \bibinfo {author} {\bibfnamefont {L.~d.}\ \bibnamefont
  {Medici}}, \bibinfo {author} {\bibfnamefont {B.}~\bibnamefont {Valenzuela}},
  \ and\ \bibinfo {author} {\bibfnamefont {E.}~\bibnamefont {Bascones}},\
  }\href {\doibase 10.1103/PhysRevB.90.115128} {\bibfield  {journal} {\bibinfo
  {journal} {Phys. Rev. B}\ }\textbf {\bibinfo {volume} {90}},\ \bibinfo
  {pages} {115128} (\bibinfo {year} {2014})}\BibitemShut {NoStop}%
\bibitem [{\citenamefont {Nakajima}\ \emph {et~al.}(2014)\citenamefont
  {Nakajima}, \citenamefont {Tanaka},\ and\ \citenamefont {Uchida}}]{naka}%
  \BibitemOpen
  \bibfield  {author} {\bibinfo {author} {\bibfnamefont {S.}~\bibnamefont
  {Nakajima}, \bibfnamefont {Masamichi~Ishida}}, \bibinfo {author}
  {\bibfnamefont {T.}~\bibnamefont {Tanaka}}, \ and\ \bibinfo {author}
  {\bibfnamefont {S.-i.}\ \bibnamefont {Uchida}},\ }\href {\doibase
  10.7566/JPSJ.83.104703} {\bibfield  {journal} {\bibinfo  {journal} {J.Phys.
  Soc. Japan}\ }\textbf {\bibinfo {volume} {83}},\ \bibinfo {pages} {104703}
  (\bibinfo {year} {2014})}\BibitemShut {NoStop}%
\bibitem [{\citenamefont {Wang}\ \emph {et~al.}(2012)\citenamefont {Wang},
  \citenamefont {Hu}, \citenamefont {Chen}, \citenamefont {Yuan}, \citenamefont
  {Li}, \citenamefont {Chen},\ and\ \citenamefont {Xiang}}]{WangNL}%
  \BibitemOpen
  \bibfield  {author} {\bibinfo {author} {\bibfnamefont {N.~L.}\ \bibnamefont
  {Wang}}, \bibinfo {author} {\bibfnamefont {W.~Z.}\ \bibnamefont {Hu}},
  \bibinfo {author} {\bibfnamefont {Z.~G.}\ \bibnamefont {Chen}}, \bibinfo
  {author} {\bibfnamefont {R.~H.}\ \bibnamefont {Yuan}}, \bibinfo {author}
  {\bibfnamefont {G.}~\bibnamefont {Li}}, \bibinfo {author} {\bibfnamefont
  {G.~F.}\ \bibnamefont {Chen}}, \ and\ \bibinfo {author} {\bibfnamefont
  {T.}~\bibnamefont {Xiang}},\ }\href {\doibase 10.1088/0953-8984/24/29/294202}
  {\bibfield  {journal} {\bibinfo  {journal} {J. Phys. Condens. Matter}\
  }\textbf {\bibinfo {volume} {24}},\ \bibinfo {pages} {294202} (\bibinfo
  {year} {2012})}\BibitemShut {NoStop}%
\bibitem [{\citenamefont {Schafgans}\ \emph {et~al.}(2012)\citenamefont
  {Schafgans}, \citenamefont {Moon}, \citenamefont {Pursley}, \citenamefont
  {LaForge}, \citenamefont {Qazilbash}, \citenamefont {Sefat}, \citenamefont
  {Mandrus}, \citenamefont {Haule}, \citenamefont {Kotliar},\ and\
  \citenamefont {Basov}}]{Schafgans2012}%
  \BibitemOpen
  \bibfield  {author} {\bibinfo {author} {\bibfnamefont {A.~A.}\ \bibnamefont
  {Schafgans}}, \bibinfo {author} {\bibfnamefont {S.~J.}\ \bibnamefont {Moon}},
  \bibinfo {author} {\bibfnamefont {B.~C.}\ \bibnamefont {Pursley}}, \bibinfo
  {author} {\bibfnamefont {A.~D.}\ \bibnamefont {LaForge}}, \bibinfo {author}
  {\bibfnamefont {M.~M.}\ \bibnamefont {Qazilbash}}, \bibinfo {author}
  {\bibfnamefont {A.~S.}\ \bibnamefont {Sefat}}, \bibinfo {author}
  {\bibfnamefont {D.}~\bibnamefont {Mandrus}}, \bibinfo {author} {\bibfnamefont
  {K.}~\bibnamefont {Haule}}, \bibinfo {author} {\bibfnamefont
  {G.}~\bibnamefont {Kotliar}}, \ and\ \bibinfo {author} {\bibfnamefont
  {D.~N.}\ \bibnamefont {Basov}},\ }\href {\doibase
  10.1103/PhysRevLett.108.147002} {\bibfield  {journal} {\bibinfo  {journal}
  {Phys. Rev. Lett.}\ }\textbf {\bibinfo {volume} {108}},\ \bibinfo {pages}
  {147002} (\bibinfo {year} {2012})}\BibitemShut {NoStop}%
\bibitem [{Note1()}]{Note1}%
  \BibitemOpen
  \bibinfo {note} {Since the high energy($>$~12\protect \tmspace +\thinmuskip
  {.1667em}000\protect \ensuremath {~\protect \textrm {cm}^{-1}}) optical
  spectral we observed does not vary with the temperatue, we do not take the
  spectral weight in this area into consideration.}\BibitemShut {Stop}%
\bibitem [{\citenamefont {Mandal}\ \emph {et~al.}(2014)\citenamefont {Mandal},
  \citenamefont {Cohen},\ and\ \citenamefont {Haule}}]{mandal}%
  \BibitemOpen
  \bibfield  {author} {\bibinfo {author} {\bibfnamefont {S.}~\bibnamefont
  {Mandal}}, \bibinfo {author} {\bibfnamefont {R.~E.}\ \bibnamefont {Cohen}}, \
  and\ \bibinfo {author} {\bibfnamefont {K.}~\bibnamefont {Haule}},\ }\href
  {\doibase 10.1103/PhysRevB.90.060501} {\bibfield  {journal} {\bibinfo
  {journal} {Phys. Rev. B}\ }\textbf {\bibinfo {volume} {90}},\ \bibinfo
  {pages} {060501} (\bibinfo {year} {2014})}\BibitemShut {NoStop}%
\bibitem [{\citenamefont {Cheng}\ \emph {et~al.}(2012)\citenamefont {Cheng},
  \citenamefont {Hu}, \citenamefont {Chen}, \citenamefont {Xu}, \citenamefont
  {Zheng}, \citenamefont {Luo},\ and\ \citenamefont {Wang}}]{Cheng2012}%
  \BibitemOpen
  \bibfield  {author} {\bibinfo {author} {\bibfnamefont {B.}~\bibnamefont
  {Cheng}}, \bibinfo {author} {\bibfnamefont {B.~F.}\ \bibnamefont {Hu}},
  \bibinfo {author} {\bibfnamefont {R.~Y.}\ \bibnamefont {Chen}}, \bibinfo
  {author} {\bibfnamefont {G.}~\bibnamefont {Xu}}, \bibinfo {author}
  {\bibfnamefont {P.}~\bibnamefont {Zheng}}, \bibinfo {author} {\bibfnamefont
  {J.~L.}\ \bibnamefont {Luo}}, \ and\ \bibinfo {author} {\bibfnamefont
  {N.~L.}\ \bibnamefont {Wang}},\ }\href {\doibase 10.1103/PhysRevB.86.134503}
  {\bibfield  {journal} {\bibinfo  {journal} {Phys. Rev. B}\ }\textbf {\bibinfo
  {volume} {86}},\ \bibinfo {pages} {134503} (\bibinfo {year}
  {2012})}\BibitemShut {NoStop}%
\bibitem [{\citenamefont {Kresse}\ and\ \citenamefont
  {Furthm\"{u}ller}(1996)}]{Kre1996}%
  \BibitemOpen
  \bibfield  {author} {\bibinfo {author} {\bibfnamefont {G.}~\bibnamefont
  {Kresse}}\ and\ \bibinfo {author} {\bibfnamefont {J.}~\bibnamefont
  {Furthm\"{u}ller}},\ }\href {\doibase 10.1016/0927-0256(96)00008-0}
  {\bibfield  {journal} {\bibinfo  {journal} {Computational Materials Science}\
  }\textbf {\bibinfo {volume} {6}},\ \bibinfo {pages} {15} (\bibinfo {year}
  {1996})}\BibitemShut {NoStop}%
\bibitem [{\citenamefont {Kresse}\ and\ \citenamefont
  {Hafner}(1993)}]{Kre1993}%
  \BibitemOpen
  \bibfield  {author} {\bibinfo {author} {\bibfnamefont {G.}~\bibnamefont
  {Kresse}}\ and\ \bibinfo {author} {\bibfnamefont {J.}~\bibnamefont
  {Hafner}},\ }\href {\doibase 10.1103/PhysRevB.47.558} {\bibfield  {journal}
  {\bibinfo  {journal} {Phys. Rev. B}\ }\textbf {\bibinfo {volume} {47}},\
  \bibinfo {pages} {558} (\bibinfo {year} {1993})}\BibitemShut {NoStop}%
\bibitem [{\citenamefont {Kresse}\ and\ \citenamefont
  {Furthm\"uller}(1996)}]{Kre}%
  \BibitemOpen
  \bibfield  {author} {\bibinfo {author} {\bibfnamefont {G.}~\bibnamefont
  {Kresse}}\ and\ \bibinfo {author} {\bibfnamefont {J.}~\bibnamefont
  {Furthm\"uller}},\ }\href {\doibase 10.1103/PhysRevB.54.11169} {\bibfield
  {journal} {\bibinfo  {journal} {Phys. Rev. B}\ }\textbf {\bibinfo {volume}
  {54}},\ \bibinfo {pages} {11169} (\bibinfo {year} {1996})}\BibitemShut
  {NoStop}%
\bibitem [{\citenamefont {Perdew}\ \emph {et~al.}(1996)\citenamefont {Perdew},
  \citenamefont {Burke},\ and\ \citenamefont {Ernzerhof}}]{Gene}%
  \BibitemOpen
  \bibfield  {author} {\bibinfo {author} {\bibfnamefont {J.~P.}\ \bibnamefont
  {Perdew}}, \bibinfo {author} {\bibfnamefont {K.}~\bibnamefont {Burke}}, \
  and\ \bibinfo {author} {\bibfnamefont {M.}~\bibnamefont {Ernzerhof}},\ }\href
  {\doibase 10.1103/PhysRevLett.77.3865} {\bibfield  {journal} {\bibinfo
  {journal} {Phys. Rev. Lett.}\ }\textbf {\bibinfo {volume} {77}},\ \bibinfo
  {pages} {3865} (\bibinfo {year} {1996})}\BibitemShut {NoStop}%
\bibitem [{\citenamefont {Yin}\ \emph {et~al.}(2011)\citenamefont {Yin},
  \citenamefont {Haule},\ and\ \citenamefont {Kotliar}}]{yin}%
  \BibitemOpen
  \bibfield  {author} {\bibinfo {author} {\bibfnamefont {Z.~P.}\ \bibnamefont
  {Yin}}, \bibinfo {author} {\bibfnamefont {K.}~\bibnamefont {Haule}}, \ and\
  \bibinfo {author} {\bibfnamefont {G.}~\bibnamefont {Kotliar}},\ }\href
  {\doibase 10.1038/NMAT3120} {\bibfield  {journal} {\bibinfo  {journal} {Nat.
  Mater.}\ }\textbf {\bibinfo {volume} {10}},\ \bibinfo {pages} {932} (\bibinfo
  {year} {2011})}\BibitemShut {NoStop}%
\bibitem [{\citenamefont {Hoffmann}\ and\ \citenamefont
  {Zheng}(1985)}]{Hoffmann1985}%
  \BibitemOpen
  \bibfield  {author} {\bibinfo {author} {\bibfnamefont {R.}~\bibnamefont
  {Hoffmann}}\ and\ \bibinfo {author} {\bibfnamefont {C.}~\bibnamefont
  {Zheng}},\ }\href {\doibase 10.1021/j100266a007} {\bibfield  {journal}
  {\bibinfo  {journal} {J. Phys. Chem}\ }\textbf {\bibinfo {volume} {89}},\
  \bibinfo {pages} {4175} (\bibinfo {year} {1985})}\BibitemShut {NoStop}%
\end{thebibliography}
%merlin.mbs apsrev4-1.bst 2010-07-25 4.21a (PWD, AO, DPC) hacked
%Control: key (0)
%Control: author (8) initials jnrlst
%Control: editor formatted (1) identically to author
%Control: production of article title (-1) disabled
%Control: page (0) single
%Control: year (1) truncated
%Control: production of eprint (0) enabled
%

\end{document}